\documentclass[sigconf,nonacm]{acmart}

\usepackage{multicol}
\usepackage{multirow}
\usepackage{colortbl}
\usepackage{subcaption}
\usepackage{amsmath}
\usepackage{hyperxmp}
\usepackage{amssymb}

\usepackage{bm}

\usepackage{dblfloatfix}

\AtBeginDocument{%
  \providecommand\BibTeX{{%
    \normalfont B\kern-0.5em{\scshape i\kern-0.25em b}\kern-0.8em\TeX}}}

\copyrightyear{2024}
\acmYear{2024}
\setcopyright{acmlicensed}\acmConference[GeoSearch'24]{3rd ACM SIGSPATIAL
International Workshop on Searching and Mining Large Collections of
Geospatial Data}{October 29-November 1 2024}{Atlanta, GA, USA}
\acmBooktitle{3rd ACM SIGSPATIAL International Workshop on Searching and
Mining Large Collections of Geospatial Data (GeoSearch'24), October
29-November 1 2024, Atlanta, GA, USA}
\acmDOI{10.1145/3681769.3698577}
\acmISBN{979-8-4007-1148-0/24/10}

\begin{document}


\title[Searching Multiword Place Names on Historical Maps]{Automatic Search of Multiword Place Names on Historical Maps}

\author{Rhett Olson, Jina Kim, Yao-Yi Chiang}
\email{[olso9295, kim01479, yaoyi]@umn.edu}
\affiliation{%
  \institution{University of Minnesota}
  \city{Minneapolis}
  \country{USA}
}

\renewcommand{\shortauthors}{Olson et al.}

\begin{abstract}
Historical maps are invaluable sources of information about the past, and scanned historical maps are increasingly accessible in online libraries. To retrieve maps from these large libraries that contain specific places of interest, previous work has applied computer vision techniques to recognize words on historical maps, enabling searches for maps that contain specific place names. However, searching for multiword place names is challenging due to complex layouts of text labels on historical maps. This paper proposes an efficient query method for searching a given multiword place name on historical maps. Using existing methods to recognize words on historical maps, we link single-word text labels into potential multiword phrases by constructing minimum spanning trees. These trees aim to link pairs of text labels that are spatially close and have similar height, angle, and capitalization. We then query these trees for the given multiword place name. We evaluate the proposed method in two experiments: 1) to evaluate the accuracy of the minimum spanning tree approach at linking multiword place names and 2) to evaluate the number and time range of maps retrieved by the query approach. The resulting maps reveal how places using multiword names have changed on a large number of maps from across history. 
\end{abstract}

\ccsdesc[500]{Information systems~Specialized information retrieval}
\ccsdesc[500]{Information systems~Spatial-temporal systems}
\ccsdesc[500]{Applied computing~Digital libraries and archives}
\ccsdesc[500]{Applied computing~Cartography \vspace{-.1in}}

\keywords{historical maps, cartography, toponym, minimum spanning tree}

\maketitle

\section{Introduction}
\begingroup
    \renewcommand{\thefootnote}{}
    \footnotetext{\textit{This is the author’s version of the work and it is posted here for personal use.}}
\endgroup
Historical maps provide unique views into the past for scholars across many disciplines~\cite{kyramargiou2020changing, sousa2010wetland}. To facilitate large-scale scholarly use of maps, recent work has aimed to make historical map documents accessible and searchable online.
With continuous efforts in digitizing historical maps, a large number of scanned historical maps are publicly available online (e.g., the David Rumsey Historical Map Collection~\footnote{\url{https://www.davidrumsey.com/}} contains roughly 60,000 scanned, georeferenced maps). Previous studies have applied computer vision techniques to automatically extract searchable text labels, consisting of recognized words from maps and bounding polygons of where the words appear~\cite{li2020automatic, mapkurator, lin2024hyper}. Leveraging these text labels, recent work has proposed a framework to automatically retrieve maps of a given place that display when all the different names for that place have been used throughout history~\cite{olson2023automatic}. 
However, one of this framework's limitations is identifying place names that consist of multiple words. For example, the framework cannot retrieve historical maps containing place names containing at least two words, such as ``United States of America'' or ``North Dakota''. This limitation is because automatically recognized text labels often contain a single word, not multiple words of an entire multiword place name, due to the complex cartographic patterns of text curvature, text orientation, font size, and word spacing used on historical maps.

\begin{figure}[h]
    \centering
    \includegraphics[width=0.85\linewidth]{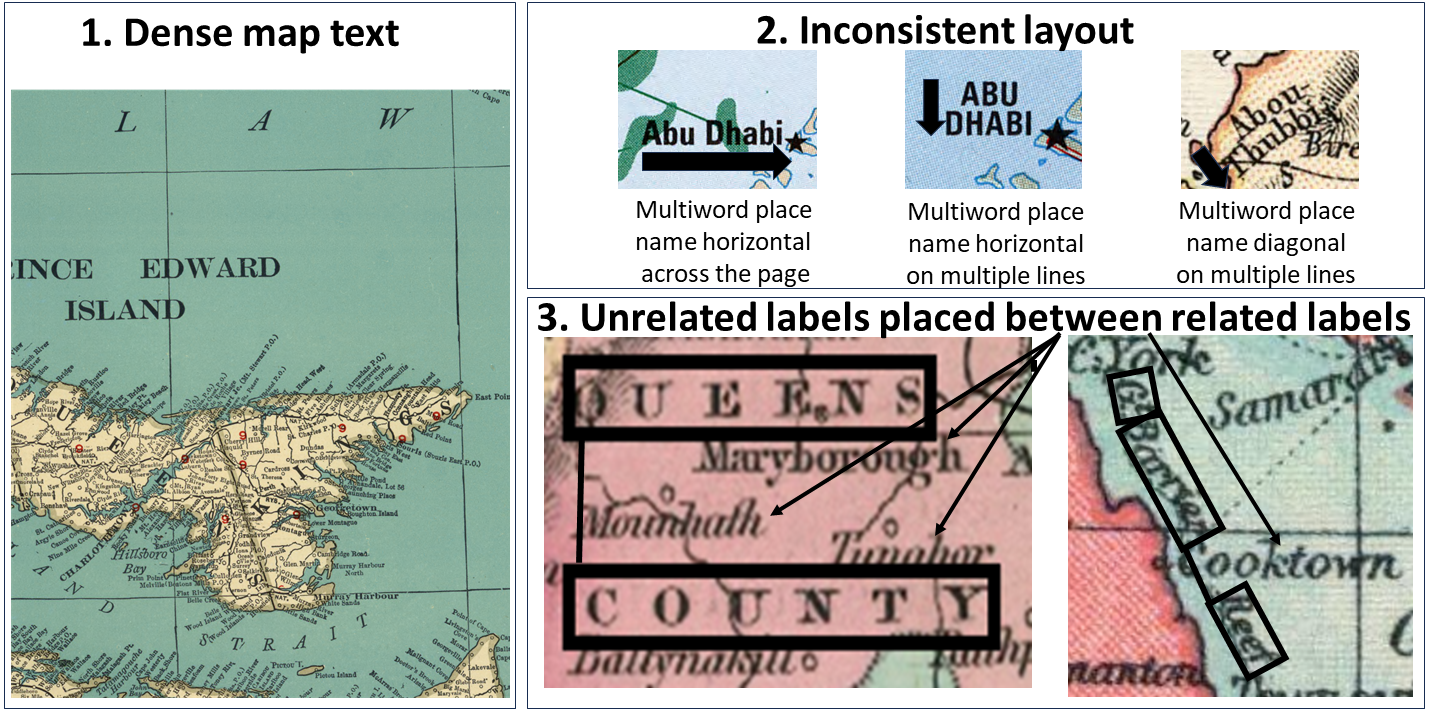}
    \vspace{-0.1in}
    \caption{Challenges of linking together multiword place names on historical maps.}
    \label{fig:linking_challenges}
    \vspace{-0.15in}
\end{figure}
A naive approach to searching for multiword place names on scanned historical maps is to search for each individual word in the multiword name among the text labels from the map, reporting a match if all words in the multiword name are found on the map. However, such an approach has a high risk of incorrectly recognizing a multiword place-name where the labels are not related on the map (e.g., recognizing the labels ``North'' and ``Dakota'' as the name of a U.S. state when in reality the label ``North'' refers to a direction on a compass, and ``Dakota'' refers to ``Dakota County''). Many multiword place names include words that are very common on maps, heightening this risk (e.g., words like ``north'', ``new'' and ''city'' appear in many multiword place names and are very common on maps). To mitigate this risk of false query results, we first need a method to link words into the most likely potential phrases.

Previous work on document layout understanding aims to link multiple words into phrases based on spatial relations of detected bounding boxes~\cite{binmakhashen2019document}. However, existing work mainly focuses on documents such as receipts or reports comprising horizontally/vertically aligned text labels. In the case of scanned historical maps, alignments of text labels are not uniform within scanned maps and across different scanned maps. Figure ~\ref{fig:linking_challenges} shows three main challenges in linking text labels of place phrases on historical maps: 1) the density of text labels on maps, 2) the inconsistency with which multiword place names are laid out, and 3) the presence of unrelated text labels appearing between text labels that are in multiword phrases.

\begin{figure*}[t]
    \centering
    \includegraphics[width=0.85\linewidth]{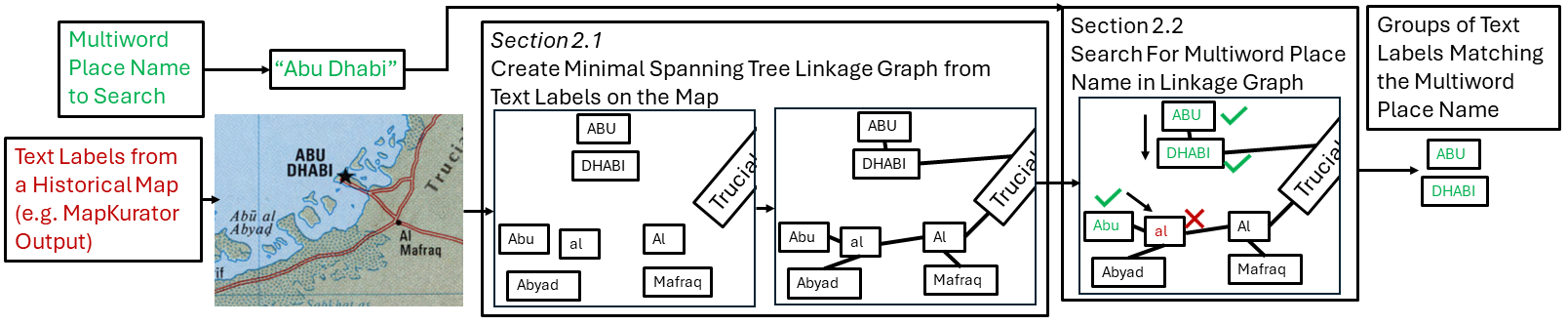}
    \vspace{-0.1in}
    \caption{An overview of our method for searching a multiword place name in a given set of text labels from a historical map.}
    \vspace{-0.2in}
    \label{fig:approach}
\end{figure*}

To overcome these challenges, we propose a simple and efficient approach for searching multiword place names among recognized text from historical maps. Figure ~\ref{fig:approach} illustrates this approach, which consists of two steps: 1) constructing a minimum spanning tree (MST), which links text labels into potential phrases, and 2) searching for the multiword place name in this tree.

\section{Methodology}

Our approach for searching for multiword place names on maps takes in two inputs. The first is a set of text labels $L = \{l_1, ..., l_n\}$ obtained from a text spotting method specialized for historical maps (e.g. mapKurator, Pallete~\cite{li2020automatic,mapkurator, lin2024hyper}). Each text label $l_i$ consists of a single word and a bounding box of the word's location on the map. The second input is a multiword place name to search. Our goal is to search for groups of text labels that combine to make the multiword place name. To avoid incorrectly returning multiword phrases composed of text labels that are not part of the same phrase on the map, we first construct a sparse graph $G = (L, E)$, where each edge $(l_i, l_j) \in E$ represents that text label $l_i$ has a high likelihood of appearing next to label $l_j$ in a multiword phrase. We then search this linkage graph for the multiword name.
\subsection{Linking Multiword Place Names}
\label{section:constructing_linkage_graphs}
To understand how likely pairs of text labels are to be adjacent in a multiword phrase, we examine explicit visual features of text labels, including distance, height (i.e., font size), orientation, and capitalization. In the most straightforward cases, text labels within a multiword place name are closer to each other than text labels that belong to other place names. Thus, we leverage the distance between text labels as an important feature. However, there are challenging cases where unrelated text labels appear between text labels that form a multiword place name. In such cases, human readers can reason that these text labels are not part of the name because they might have a different font size, angle of orientation, or capitalization to the text labels in the multiword place name. For example, in Figure~\ref{fig:linking_challenges}, the unrelated text labels appearing between the multiword name ``QUEENS COUNTY'' have a smaller font size and are not in all capitals,  and the text label between the multiword name ``Gt. Barrier Reef'' is oriented at an angle nearly perpendicular to the text labels in that multiword name. From these observations, we define a heuristic called an edge cost function, which represents how unlikely two text labels $l_i$ and $l_j$ are to be adjacent in a multiword phrase. This edge cost function is based on four explicit visual features of the text labels. The first is the minimum spatial distance between the bounding boxes of the two text labels on the map image, denoted as $d(l_i, l_j)$. The second is the ratio of the height of the label with the taller bounding box to that of the label with the shorter bounding box, minus $1$. This feature, denoted as $h(l_i, l_j)$, will be minimized at $0$ for labels with the same height and, by extension, similar font sizes. The third feature, denoted as $a(l_i, l_j)$, is the sine of an angle difference between the axes of two bounding boxes, which will range from $0$ to $1$ and be minimized when the axes of the bounding boxes of two labels are aligned. The final feature, denoted as $c(l_i, l_j)$, represents the difference in capitalization between the labels. The value of $c(l_i,l_j)$ is $0$ when either the two labels are all uppercase, or neither is all uppercase, or $1$ if one is all uppercase and the other is not. 

The edge cost function, incorporating these four features, is defined as:
\begin{equation}
    cost(l_i, l_j) = d(l_i,l_j) * (1 + h(l_i,l_j)) * (1 + a(l_i,l_j)) * (1 + c(l_i, l_j))
    \label{equation:edge_cost}
\end{equation} 

The intuition behind this edge cost function is to use discrepancies in height, angle, and capitalization as separate multipliers for the distance between the text labels, where each multiplier increases above $1$ the larger the discrepancy is. The distance between text labels that are part of the same phrase varies from map to map and for text labels with different font sizes. Because of this, we use the other features as multipliers of the distance to ensure that the penalties for discrepancies in height, angle, and capitalization scale with the distances between text labels on different maps.

Many pairs of text labels could be connected in our linkage graph, but the vast majority of these pairs are not next to each other in a multiword phrase. To filter out most of these pairs, we construct a sparse graph that links only $O(n)$ out of the $O(n^2)$ possible pairs of text labels, where $n$ is the number given of text labels. We apply the edge cost function to compute the edge cost between each pair of text labels. From these edge costs, we utilize Prim's algorithm~\cite{prim1957shortest} to construct a minimum spanning tree (MST) of the text labels, which minimizes the sum of the costs of edges in the tree. Because these edge costs represent how unlikely the text labels are to be adjacent in a multiword phrase, minimizing the total edge cost aims to choose edges that are most likely to be next to each other in multiword phrases. Because MSTs include exactly $n - 1$ edges, each label will be connected to an average of less than two other text labels. This ensures that, although some of the edges in the MST may not connect labels that are part of the same phrase, there is a small chance that a text label will be incorrectly linked to another label containing another word from the queried multiword place name. For example, if searching for the name ``North Dakota'' on a map that contains a text label ``Dakota'' from a different place name (e.g. ``Dakota County''), there is a small chance that one of the few labels ``Dakota'' is connected to in an MST will have the word ``North''. 
Research on the related problem of text line grouping in scene images has also applied MSTs~\cite{pan2009text}.

\vspace{-0.1in}
\subsection{Searching for Multiword Place Names within the Linkage Graph}
\label{section:searching_msts}
To find a given multiword place name, we search the MST linkage graph for connected sequences of text labels on the map that match that name. We search for the first word of the multiword name and search for the rest of the words in the name through the edges of matching labels in the linkage graph. For example, if searching for the multiword place name ``Sault Ste. Marie'', we first search for the word ``Sault'', and for each text label whose text matches this word, we search for text labels connected to that label in the linkage graph that match the text ``Ste.'', and for each matching label we search among its neighbors for the last word, ``Marie.'' If such a path of connected text labels matching the given multiword place name is found, that group of text labels is returned by the query, and the map is considered to contain the given multiword name. If any such path exists in the linkage graph, that sequence of text labels will be found by the proposed search method in $O(n)$ time because there are $n$ text labels and $n - 1$ edges in the graph. The few edges we search are chosen by the MST, such that the edges are most likely to connect words that appear next to each other in a multiword place name, according to the edge cost function.

\section{Experiments}
\paragraph{\textbf{Evaluating Linkage Graphs}}
To evaluate the effectiveness of the MST method described in Section ~\ref{section:constructing_linkage_graphs} at linking multiword place names, we utilize a competition dataset from ICDAR 2024 containing scanned map images with annotated words and phrases~\cite{icdar2024HistoricalMaps}. We compare our method to two baselines, one that does not use MSTs and one that uses MSTs with a conventional edge cost function.

\paragraph{\textbf{Dataset}}
The competition dataset from ICDAR 2024~\cite{icdar2024HistoricalMaps} consists of 940 different $2000 \times 2000$ pixel tiles of scanned historical maps from the David Rumsey Historical Map collection. Each map tile has annotated phrases, where each phrase is represented as a list of one or more text labels, with each text label consisting of a word and its corresponding bounding polygon. To convert these bounding polygons into rectangular bounding boxes, which are needed to compute the heights and angles of text labels used in Equation ~\ref{equation:edge_cost}, we compute minimum area rectangles for each polygon.

\paragraph{\textbf{Linkage Methods}}
Our linkage method, described in Section ~\ref{section:constructing_linkage_graphs}, employs MSTs with Equation ~\ref{equation:edge_cost} as the edge cost function. We employ two baseline methods as follows. First, we employ the character distance threshold procedure used by the David Rumsey map library~\footnote{\url{https://machines-reading-maps.github.io/rumsey/\#search-for-multiple-words}}, which draws edges between every pair of labels whose bounding polygons are within the width of two characters of each other, where the width of a character is given by the width of the text label's bounding polygon divided by the length of the word on the label. The other baseline approach utilizes MSTs constructed from a different edge cost function, the Mahalanobis distance~\cite{yin2009handwritten, kaya2019deep}. The Mahalanobis distance between text labels $l_i$ and $l_j$ is given by: 
\begin{equation}
MahalanobisDistance(l_i, l_j) = \sqrt{(l_i - l_j)^T M (l_i - l_j)},
\label{equation:mahalanobis_distance}
\end{equation}
where $(l_i - l_j) \in \mathbb{R}^4$ is defined as $(d(l_i,l_j), h(l_i,l_j), a(l_i,l_j), c(l_i, l_j))$. $M \in \mathbb{R}^{4 \times 4}$ is a positive semi-definite matrix learned from annotated data by solving a convex programming problem that aims to find the matrix $M$ that minimizes the sum of the squared Mahalanobis distance between text labels next to each other in a multiword phrase under the constraint that the sum of the squared Mahalanobis distance between unrelated text labels is at least $1$.


We conduct a 5-fold cross-validation on the dataset of 940 annotated map tiles. The Mahalanobis distance baseline requires a training set, whereas our proposed method and the baseline using a character distance threshold do not. Therefore, we use the training set from each fold only for the Mahalanobis distance baseline. The test set from each fold is then used to evaluate all three methods.

As evaluation metrics for these linkage graphs, we use recall and precision. We compute recall as
\begin{equation}
    \frac{\text{\# correctly linked multiword phrases}}{\text{total \# of multiword phrases}}
\end{equation}
We compute precision as 
\begin{equation}
    \frac{\text{\# edges that correctly link all words in multiword phrases}}{\text{total \# of edges in linkage graphs}}
\end{equation}

Note that these definitions of precision and recall differ from the conventional definitions. In our definition of precision, true positives are defined based on the successful connections between words in multiword phrases rather than simply the number of relevant items retrieved.

%


\begin{table}[h]
\centering
\scalebox{0.8}{
\begin{tabular}{l|cc}
Linkage Method & Precision (\%) & Recall (\%)\\ \midrule
\begin{tabular}[c]{@{}l@{}}Character Distance\\ Threshold\end{tabular} & 25.33 & 82.28 \\ \midrule
\begin{tabular}[c]{@{}l@{}}Mahalanaobis\\ Distance MST\end{tabular}    & 25.16 & 79.68 \\ \midrule
\textbf{Our Method} & \textbf{26.15} & \textbf{82.61}
\end{tabular}}     
\caption{Evaluation of three linkage graph methods on the ICDAR 2024 map text competition dataset~\cite{icdar2024HistoricalMaps}.}
\vspace{-0.1in}
\label{table:linkage_results_table}
\end{table}

\begin{figure}[h]
    \centering
    \includegraphics[width=0.8\linewidth]{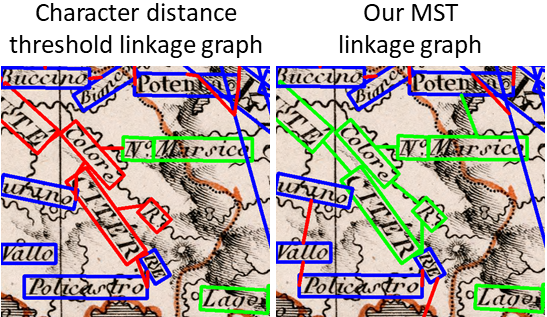}
    \caption{Visualization of linkage graphs constructed from map text labels the character distance threshold baseline (left) and our method (right).  The colors of the bounding boxes indicate:
    \textcolor{green}{$\square$}: correctly linked multiword phrase
    \textcolor{blue}{$\square$}: single word \textcolor{red}{$\square$}: incorrectly linked multiword phrase.
}\label{fig:mst_visualization}
    \vspace{-.2in}
\end{figure}

\paragraph{\textbf{Linkage Results}}
Figure~\ref{fig:mst_visualization} shows a visual example comparing the effectiveness of the constructed linkage graphs. Table~\ref{table:linkage_results_table} reports the average precision and recall of the 5-fold cross-validation. The results show that our method performs slightly higher for all evaluation metrics than both baseline methods. Compared with the character distance threshold, the highest-performing baseline method, our method achieves $82.61\%$ ($+0.33\%$) recall and $26.15\%$ ($+0.82\%$) precision. 

The recall of $82.61\%$ shows our method correctly links and enables searches for a large majority of the multiword place names on historical maps. As described in Section~\ref{section:constructing_linkage_graphs}, our MST approach filters out the vast majority of possible incorrect edges by creating a sparse tree graph. However, the precision of $26.15\%$ indicates that linkage graphs have many incorrect edges that link text labels that are not adjacent in multiword phrases. These incorrect edges can cause incorrect query results if the words on the text labels they connect both appear in the queried place name. We further plan to improve the linkage graph method to mitigate the low precision.


\vspace{-0.1in}
\paragraph{\textbf{Searching for Multiword Place Names}}
To demonstrate the effectiveness of our approach at finding multiword place names on maps, we applied our method for multiword place name search to the experimental time-sequenced map query system presented in~\cite{olson2023automatic}. This query system searches through over $100$ million text labels from roughly $60,000$ historical maps for appearances of a given place, searching for all name variants for that place, according to a gazetteer. For example, a query for ``Sault Ste. Marie'' will also search for other names for that place including ``St. Mary's Falls'' and ``Fort Brady''. Applying our MST phrase linking and searching method enables these queries to find multiword place names.

\vspace{-0.1in}
\paragraph{\textbf{Query Samples}}
To evaluate the effectiveness of the query approach in recognizing multiword place names, we process queries for the 15 most populous cities in the world that currently use multiword place names. 
We obtain these names from a dataset of roughly 47,000 cities ~\footnote{\url{https://simplemaps.com/data/world-cities}} by filtering for cities whose names contain at least one space character (i.e. multiple words), and selecting the 15 most populous results. In addition, we run queries on 49 countries that use multiword names in English, according to OpenStreetMap,~\footnote{\url{https://wiki.openstreetmap.org/wiki/Countries_of_the_world}} and the 20 largest lakes in the world that use multiword names according to NASA JPL.~\footnote{ \url{https://largelakes.jpl.nasa.gov/world-lakes}}

\vspace{-0.1in}
\paragraph{\textbf{Query Results}}
\begin{figure}[h]
    \centering
    \vspace{-0.1in}
    \includegraphics[width=1\linewidth]{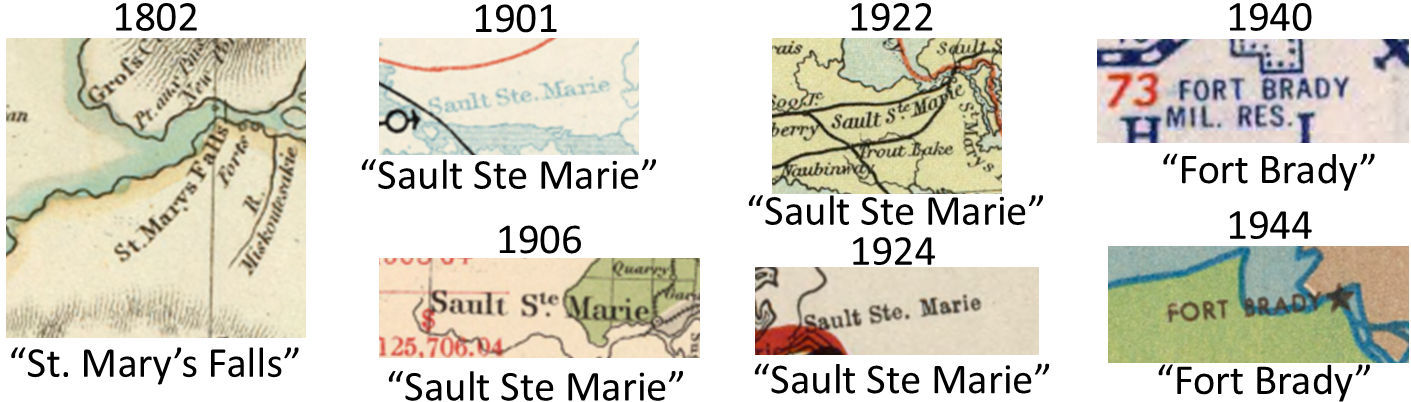}
    \vspace{-0.1in}
    \caption{Maps sampled from a time-sequenced map query for ``Sault Ste. Marie''.}
    \label{fig:query_results}
    \vspace{-0.1in}
\end{figure}
\begin{table}[h]
\scalebox{0.83}{
\begin{tabular}{l|cc}
\begin{tabular}[c]{@{}l@{}}Multiword Place Name\\ Query Sample \end{tabular}                                                               & \begin{tabular}[c]{@{}l@{}}Avg. \# of Retrieved Maps\\ with Multiword Names\end{tabular} & \begin{tabular}[c]{@{}l@{}}Avg. Time\\Span of Maps \end{tabular} \\ \midrule
49 Countries                                                               & 39.89 & 176 years \\ \midrule
20 Largest Lakes                                                                & 130.26 & 233 years \\ \midrule
\begin{tabular}[c]{@{}l@{}}15 Most\\ Populous Cities\end{tabular} &  145.26 & 265 years         
\end{tabular}}
\caption{Query results for three different samples of places that use multiple word names.}\label{table:query_results}
\vspace{-0.1in}
\end{table}
Table ~\ref{table:query_results} shows statistics of the time-sequenced map query results. These queries retrieved an average of 80.22 maps matching multiword place names per query, and these maps spanned an average time range of 205  years. Figure ~\ref{fig:query_results} shows maps sampled from the 42 time-sequenced map results of a query for ``Sault Ste. Marie''. These results show that this approach can be used to retrieve many historical maps of places that use multiword names, revealing changes in each place across wide ranges of history.

\section{Conclusion}
We propose an approach to searching for multiword place names among single-word text labels on historical maps. The proposed approach links words into potential phrases by constructing minimum spanning trees (MSTs) that minimize the differences in location, text height, orientation angle, and capitalization among linked text labels. Experimental queries show the potential for this approach to retrieve many maps including specific multiword place names, which span hundreds of years of history. Experimental evaluation of the MST linkages shows that our query method correctly links over $82\%$ of ground truth multiword place names from an annotated dataset of maps but also links many pairs of text labels that are not part of the same phrase. We plan to improve these queries by investigating methods that link text labels with higher precision.    

\vspace{-.1in}
\begin{acks}
We thank David and Abby Rumsey for their generous support and for making scanned maps publicly available. All scanned maps used in figures are credited to the David Rumsey Map Collection, David Rumsey Map Center, Stanford Libraries. This project was supported by the University of Minnesota's Office of Undergraduate Research.
\end{acks}

\vspace{-.1in}
\bibliographystyle{ACM-Reference-Format}
\bibliography{biblio}


\end{document}